# Impairment of insulin-stimulated glucose utilization is associated with burn-induced insulin resistance in mouse muscle by hyperinsulinemic-isoglycemic clamp


Takeshi Yamagiwa,[1,2,3] Yong-Ming Yu,[1,2,3] Yoshitaka Inoue,[1,2,3] Vasily V. Belov,[3, 4]

Mikhail I. Papisov,[3,4] Sadaki Inokuchi,[6] Masao Kaneki,[2,3,7] Morris F. White,[3,5]

Alan J. Fischman,[2] and Ronald G. Tompkins[1,3]

[1] Department of Surgery, Massachusetts General Hospital, Boston, Massachusetts, USA;

[2] Shriners Hospitals for Children, Boston, Massachusetts, USA;

[3] Harvard Medical School, Boston, Massachusetts, USA;

[4] Department of Radiology, Massachusetts General Hospital, Boston, Massachusetts, USA;

[5] Department of Medicine, Boston Children's Hospital, Boston, Massachusetts, USA

[6] Department of Emergency and Critical Care Medicine, Tokai University School of Medicine, Isehara, Kanagawa, Japan

[7] Department of Anesthesia, Critical Care and Pain Medicine, Massachusetts General Hospital, Boston, Massachusetts, USA







**Running Head:** Impaired glucose utilization and insulin resistance

**Corresponding Author:**

Ronald G. Tompkins, M.D., Sc.D.

Harvard Medical School

Director, Center for Surgery, Innovation & Bioengineering

Massachusetts General Hospital, 55 Fruit St., GRB 1302

Boston, MA 02114

Email: rtompkins@mgh.harvard.edu

Phone: 617-726-3447

Fax: 617-643-4443







**ABSTRACT**

Burn-induced insulin resistance is associated with increased morbidity and mortality; however, the impact of burn injury on tissue-specific insulin sensitivity and its molecular mechanisms with consideration of insulin state remains unknown in rodent models. This study was designed to characterize a burn mouse model with tissue-specific insulin resistance under insulin clamp conditions. C57BL6/J mice were subjected to 30% full-thickness burn injury and underwent the combination of hyperinsulinemic isoglycemicclamp (HIC) and positron emission tomography (PET). Hepatic glucose production (HGP) and peripheral glucose disappearance rate (Rd) were measured at different time points up to 7 days post injury. Burned mice showed a significant fasting hypoglycemia and hypoinsulinemia ($P < 0.01$) on post-burn day (PBD) 3 and 7 along with significantly higher energy expenditure ($P < 0.01$). HICon PBD 3 demonstrated that burn injury induced systemic insulin resistance, resulting from a significant decrease in insulin-stimulated Rd ($33.0 \pm 10.2$ vs $68.3 \pm 5.9$ mg/kg/min; $P < 0.05$). In contrast, HGP of burned and sham mice was comparable both in the basal and clamp period. PET on PBD 3 showed a lower insulin-stimulated 18F-labeled 2-fluoro-2-deoxy-D-glucose uptake in the quadriceps of burned mice compared with sham-burned mice. Gastrocnemius muscle harvested from burned mice on PBD 3





showed decreased insulin-stimulated tyrosine phosphorylation of insulin receptor substrate-1 to 34.7% of that in sham-burn mice by immunoblotting analysis ($P < 0.05$). These findings suggest that impaired insulin-stimulated Rd in skeletal muscle, not elevated HGP, plays a role in the development of burn-induced insulin resistance in a mouse model.

## INTRODUCTION

Critically ill patients including those with burn injury develop insulin resistance and hyperglycemia, which lead to poor wound healing(30), skeletal muscle wasting (26), and increased mortality(10, 25). Intensive insulin therapy has been shown to improve glucose homeostasis in burned children(9); however, the increased incidence of hypoglycemic episodes with negative outcomes(1, 4, 17) requires a better understanding of the underlying molecular mechanism(s) of the burn-induced insulin resistance. This new knowledge will likely contribute to improved patient outcomes and more targeted approaches to therapy.

It is well recognized that acute insulin resistance is complex and tissue specific in liver, skeletal muscle, and adipose tissue(19-21). Studies investigating the potential molecular mechanisms for burn induced insulin resistance have demonstrated that burn injury impairs the insulin receptor substrate (IRS)-1/ phosphatidyl- inositol 3-kinase (PI3K)/ Akt signal pathway in liver(3) and skeletal muscle(13, 14, 28). These findings, however, were obtained from isolated tissues from rodent burn models after differing doses of insulin challenge and only whole-body glucose metabolism was assessed, by glucose tolerance test (GTT) and/or insulin tolerance test (ITT). We believe that the investigations exploring molecular mechanism(s) should be conducted







on appropriate tissue sample which reflects an accurate tissue-specific insulin resistance in vivo evaluated by well-characterized animal models. Recently, we reported that burn injury was associated with a 49% increase in hepatic glucose production (HGP) in the fasting state with post-burn hyperglycemia by the hyperinsulinemic-euglycemic clamp technique in a rabbit model(31). The insulin clamp technique is the gold standard for quantitative assessment of insulin sensitivity(6). Ininsulin clamp technique, whole-body insulin action is qualified as the glucose infusion rate (GIR) required to maintain euglycemia or isoglycemia in response to a constant rate of insulin infusion. Additionally, HGP and the rate of peripheral glucose disappearance (Rd) can be obtained by infusing radioisotope glucose. Furthermore, positron emission tomography (PET), a quantitative imaging technique, allows the assessment of glucose dynamics in individual tissues of living animals. The combination of insulin clamp and PET techniques allows for the evaluation of glucose metabolism in individual tissues. Although these approaches are technically challenging to perform in small animals, a well-characterized burn mouse model has been utilized in the development of genetically modified animals, with tissue-specific knockout of the insulin receptor(2, 16). The purpose of the current study was to characterize both the impact of burn injury on tissue-specific insulin sensitivity and the molecular basis of insulin resistance under







insulin clamp conditions in a mouse model utilizing the combination of the insulin

clamp technique and PET.







## MATERIALS AND METHODS

*Animals.* All procedures were approved by the Institutional Animal Care Committee

(IACUC) at Massachusetts General Hospital. The animal care facility is accredited by

the Association for Assessment and Accreditation of Laboratory Animal Care

(AAALAC). Male 10 to 12-week-old C57BL/6 mice (Charles River Laboratories;

Wilmington, MA) were housed in a pathogen-free animal facility at Shriners Hospital

under controlled temperature, humidity, and a 12:12h light-dark cycle. The mice were

provided with standard rodent chow and water ad libitum. A full-thickness third-degree

burn injury comprising 30% of total body surface area was produced by immerging the

abdomen for 6 sec and both flanks for 4 sec in 80°C water under general anesthesia with

ketamine (100 mg/kg BW, IP) and xylazine (10 mg/kg BW, IP). We confirmed that this

procedure produced full-thickness burn injury by light microscopy of hematoxylin and

eosin stained sections. Sham-burn animals were treated in the same manner with the

exception that these animals were immerged in room temperature water.   Immediately

after burn or sham-burn, all animals received fluid resuscitation with normal saline (40

ml/kg BW, IP). Buprenorphine (0.1 mg/kg BW, IM) was administered every 12 h for up

to 72 h after the injury. This study was conducted in pair-fed burn and sham-burned

mice.



                                                 

*Indirect calorimetric analysis.* At 1, 3 and 7 days after burn or sham-burn, indirect

gas calorimetry (TSE Systems, Germany) was used to evaluate respiratory exchange

ratio (RER), energy expenditure (EE), oxygen consumption ($VO_2$), and resting carbon

dioxide production ($VCO_2$). These parameters were calculated automatically by a

computer system based on body weight. The mice were placed into a temperature- and

humidity-controlled metabolic chamber for 24 h and had ad libitum access to water and

food.

*Glucose tolerance test (GTT).* At 1, 3, 7 days after burn or sham-burn and following

15-h of fasting, 2.0 g/kg of glucose was injected intraperitoneally. Blood glucose levels

and plasma insulin concentration were measured by sampling blood from the tail vein

using a FreeStyle Lite glucometer (Abbott, Almeda, CA) and ELISA kit (CRYSTAL

CHEM INC, Downers Grove, IN) before and at 15, 30, 60, 90, 120 min after glucose

injection.

*Insulin tolerance test (ITT).* At 1, 3, 7 days after burn or sham-burn, following 5-h

of fasting, mice were injected into the abdomen with insulin (1.0 U/kg BW; Humulin R,







Eli Lilly, dianapolis, IN). Blood glucose levels obtained from the tail vein were recorded for 120 min after the insulin injection.

*Hyperinsulinemic-sisoglycemic clamp (HIC) in a conscious mouse.* We performed the HIC technique, which maintained blood glucose at fasting levels, because basal blood glucose level differed significantly between burn and sham-burn mice in preliminary experiments and large acute changes in glycemia may alter insulin sensitivity(22). Four days before burn or sham-burn, the mice were anesthetized with 2.0% of isoflurane and a polyurethane catheter (0.025" outer diameter) was inserted into the right jugular vein for infusions. Animals were housed individually after the surgery, and their body weight (BW) and food consumption were recorded daily.

At 1, 3, 7 days after burn or sham-burn, following 13-h fasting, the mice were placed into a rat-sized restrainer (2.5" diameter and 8.5" long) with tail-tethered for blood sampling. After 2-h acclimatization, baseline blood glucose and plasma insulin concentrations were determined. The protocol consisted of a 90-min tracer equilibration period (t = -90 min to 0 min) and a 120-min clamp period (t = 0 min to 120 min). A 600 μmol/kg bolus injection of [U-13C6] glucose (Cambridge Isotope Laboratories, Andover, MA) was administered at t = -90 min, followed by a 7.5 μmol/





kg/min infusion for 90 min. The HIC was initiated at t = 0 min with a bolus insulin injection (50 mU/kg BW) for 3 min followed by continuous insulin infusion (2.5 mU/kg BW) for 120 min. Blood glucose concentration was determined every 10 min to adjust the infusion rate of 50% dextrose (Hospira, Lake Forest, IL) to maintain the target glucose level at baseline concentration ± 10 mg/dl. Steady state of the clamp was defined as a 30-min or longer period during which the coefficient of variation for GIR and blood glucose were less than 5%. Blood samples (40 - 60 µl) were taken every 15 min from t = 75 min to determine [U-13C6] glucose enrichment. The clamp plasma insulin concentrations were determined from the sample taken at t = 120 min. The enrichments of [U-13C6] glucose were measured by gas chromatography mass spectrometry (GC-MS, Hewlett Packard 5985B GC/MS System, Palo Alto, CA) as described previously(31). Immediatly after HIC, the mice were anesthetized with sodium pentobarbital (100 mg/kg BW, IP), and gastrocnemius muscle was excised. The tissue samples were snap-frozen and kept –80° C until the biochemical analysis.

*Calculations.*    Whole body glucose turn over (Ra) was determined with the steady-state isotope tracer dilution approach as described previously (26): Ra = $Q_{iso}$ ($E_{iso}$/ Ep – 1), where $Q_{iso}$ is the infusion rate of [U-13C6] glucose, $E_{iso}$ is the isotope







enrichment of [U-13C6] glucose in the infusate, and $E_p$ is the plateau level of [U-13C6]

glucose enrichment. HGP was calculated by subtracting the GIR from Ra. The glucose

clamp-derived index of insulin sensitivity ($SI_{clamp}$) was defined as M/(G x $\Delta$I), where M

is the steady state of GIR, G is the steady state blood glucose concentration, and $\Delta$I is

the difference between basal and steady state plasma insulin concentrations(15).

Spearman rank-coefficient of correlation value between GIR at steady state and $SI_{clamp}$

was measured.

*MicroPET.* We prepared a separate group of mice in the same manner as the HIC

animals for microPET study. At 3 days after burn or sham-burn injury, microPET was

conducted at fasting (basal) and steady state condition (clamp) to investigate individual

muscle glucose metabolism at different conditions. Following 30-min steady state with

HIC (clamp), or without HIC (basal), the mice were transferred to the gantry of a

microPET Focus 220 camera (Siemens Medical Solutions, Knoxville, TN) and

anesthetized with 1.0% isoflurane. Blood glucose concentration was controlled at

fasting level throughout the study. 1.2mCi of 18F-labeled 2-fluoro-2-deoxy-D-glucose

([18]FDG) obtained from PETNET Solutions (Woburn, MA) was injected intravenously

through the catheter with immediate start of dynamic PET image acquisition. The







dynamic acquisition with 1 min frames was performed for 25 min followed by 5 min static acquisition recorded at 30 min post administration. Imaging was carried out using a custom PET/CT imaging system consisting of MicroPET Focus 220 PET scanner (Siemens Medical Solutions, TN) and CereTom NL 3000 CT scanner (Neurologica, MA, USA).

Five circular regions of interest were manually drawn inside selected tissue, and the mean activity concentration was corrected for radioactive decay, and expressed as standardized uptake value (SUV).

*Tissue homogenization and immunoblotting.* Gastrocnemius muscle excised from the 15h-fasted mice (basal) or mice that underwent HEC (clamp) were homogenized using a electric homogenizer (Cole-Parmer, Vernon Hills, IL) in ice-cold homogenization buffer containing 20 mM Tris-HCl (pH 7.5), 150 mM NaCl, 1 mM $Na_2EDTA$, 1 mM EGTA, 1% Triton, 2.5 mM sodium pyrophosphate, 1 mM beta-glycerophosphate, 1 mM $Na_3VO_4$, 1 µg/ml leupeptin (Cell Signaling, Danvers, MA) and 1 mM Phenylmethanesulfonyl fluoride (Sigma, St. Louis, MO), followed by centrifuging at 14,000 $g$ for 10 min at 4°C. Equal amounts of protein were subjected to SDS-polyacrylamide gel electrophoresis and were transferred to 0.2-µm nitrocellulose







membranes (BioRad, Herculs, CA). The membranes were blocked with Odyssey

blocking buffer (Odyssey, Lincoln, NE) for 1h at room temperature, followed by

incubation overnight at $4^oC$ with anti-glyceraldehyde 3- phosphate dehydrogenase

(GAPDH) (#2118),anti-Akt (#4691), anti-phospho-Akt at serine 473 (#4060),

anti-IRS-1 (#2390) (Cell Signaling, Danvers, MA), and anti-phospho IRS-1 at tyrosine

612 (#44-816G) (Life Technologies, Grand Island, NY). The membranes were then

incubated with goat anti-rabbit IRDye 800 (LI-COR, Lincoln, NE) for 1h at room

temperature, and were scanned with an Odyssey Infrared Imaging System (LI-COR,

Lincoln, NE). The images were analyzed using the Odyssey Application Software,

version 4.0 (LIL-COR, Lincoln, NE) to obtain integrated intensities. The obtained

intensities were normarized by GAPDH.

*Statistical analysis.* Statistical analysis was performed using GraphPad Prism

software (GraphPad Software, San Diego, CA). The data were compared with unpaired

t-test unless otherwise noted and repeated measure based parameters (such as BW, and

GIR) were analyzed using two-way ANOVA followed by the Bonferroni post hoc

multiple comparisons test. A value of $P < 0.05$ was considered statistically significant.

All results were expressed as mean $\pm$ SE.







## RESULTS

*Hypermetabolism, catabolism, and unexpected hypoglycemia*

Animals with 30% TBSA burn injury demonstrated a significant decrease in BW on and after PBD 3 compared with sham-burned animals (**Fig. 1A**). The weight of gastrocnemius muscle and epididymal adipose tissue in burned mice was significantly decreased on PBD 7 (Gastrocnemius; $0.6 \pm 0.02$ vs $0.4 \pm 0.03$% BW, P < 0.05, Epididymal; $1.6 \pm 0.2$ vs $0.9 \pm 0.1$% BW, P < 0.01) (**Fig. 1B**). In contrast, the liver weight of burned mice was significantly increased on PBD 7 ($6.0 \pm 0.3$ vs $5.0 \pm 0.1$% BW, P < 0.05) (**Fig. 1B**). The indirect calorimetry study demonstrated that burn injury significantly increased EE on PBD 3 ($19.3 \pm 2.3$ vs $14.1 \pm 2.3$ kcal/kg•h, P < 0.01) and 7 ($18.1 \pm 2.4$ vs $13.3 \pm 1.8$ kcal/kg•h, P < 0.01), along with increased $VO_2$, and $VCO_2$ (**Fig. 1C-G**). These findings indicate that our animal model of burn injury is consistent with the clinical course of burn patients with respect to hypermetabolism and catabolism. Blood sampling after 15-h fasting, however, demonstrated significant hypoglycemia and hypoinsulinemia in the burned mice on PBD 3 and 7, which resulted in significantly lower Homeostasis Model Assessment of Insulin Resistance (HOMA-IR, estimate of steady state beta cell function and insulin sensitivity as percentages of a normal reference population) in burned mice on PBD 3 and 7 compared






with sham-burn mice (**Table 1**).

*Systemic insulin resistance on post burn day 3 by indirect measurements*

We performed indirect measurements including GTT and ITT at various time points
to assess the impact of burn injury on systemic glucose tolerance and insulin sensitivity
in this animal model.    Areas under the curve (AUCs) of blood glucose (**Fig. 2A-C**) and
plasma insulin concentration (**Fig. 2D-F**) after GTT, calculated by the trapezoidal
formula, were comparable between burn and sham-burn mice throughout the study
period.    In contrast, ITT (**Fig. 2G-I**) showed a significantly higher AUC of blood
glucose on PBD 3 compared with sham-burn mice ($8569 \pm 353$ vs $6625 \pm 213$ mg/dL, P
$< 0.01$) (**Fig. 2H**). These findings suggested that this animal model induced whole-body
insulin resistance on PBD 3 although the animals did not develop glucose intolerance.

*Decreased insulin-stimulated glucose utilization in perioheral tissues by*

*hyperinsulinemic isoglycemic clamp*

We conducted HICat various time points to validate the ITT results and determine
whether HGP and/or peripheral Rd contributed to burn-induced insulin resistance.
Blood glucose level during the clamp period was maintained at the target range both in







burn and sham-burn mice at each time point (**Fig. 3A-C**). However, burned mice on

PBD 3 showed a significantly lower GIR than that of sham-burn mice (AUC GIR: 3383

± 604 vs 7450 ± 503 mg/kg/min, P < 0.01) (**Fig. 3E**), which indicates that burn injury

induces systemic insulin resistance on PBD 3.    There were no significant differences in

AUCs of GIR between burn and sham-burn mice on PBD 1 and 7 (Fig. 3D and F).

These results were in agreement with the results using $SI_{clamp}$, which is recognized as a

more appropriate method than GIR for the isoglycemic clamp technique because it

normalizes for differences in fasting glucose levels (22). $SI_{clamp}$ of burned mice on PBD

3 was also significantly lower than that of sham-burn mice (0.58 ± 0.23 vs 1.16 ± 0.14,

P < 0.05), and $SI_{clamp}$ showed a significant positive correlation with GIR at steady state

($r^2$ = 0.56, P < 0.01) (**Fig. 3G**).

HGP and peripheral Rd were determined by gas chromatography mass spectrometry.

HGP on PBD 3 showed no significant difference during the basal and clamp period

(**Fig. 3H**). In contrast, insulin stimulated-Rd of burned mice on PBD 3 was significantly

lower than in sham-burn mice (33.0 ± 10.2 vs 68.3 ± 5.9 mg/kg/min, P < 0.05) (**Fig. 3I**).

Additionally, plasma insulin concentration during the clamp period was maintained at a

comparable level between burned and sham-burned animals at each time point (**Fig. 3J**).

The results showed that impaired insulin-stimulated Rd in peripheral tissue acompanied







with unelevatedHGP plays an important role in developing the insulin resistance of burned mice under insulin clamp conditions.

*Decreased insulin-stimulated glucose utilization in skeletal muscle by hyperinsulinemic isoglycemic clamp and microPET*

We conducted microPET following HIC on PBD 3 to further investigate the glucose dynamics in the individual tissues.   MicroPET in the basal period showed that [18]FDG uptake in the quadriceps was comparable between burn and sham-burn mice; however, during the clamp period, insulin infusion promoted [18]FDG uptake in the quadriceps of sham-burned mice more than in burned mice (**Fig. 4A**). Quantitative analysis of SUV showed that insulin-stimulated SUV in the right quadriceps of burned mice was significantly lower than that in sham-burn mice (**Fig. 4B**). These results are further supported in the data to follow showing that burn injury impairs insulin-stimulated glucose disappearance in the skeletal muscle, and this impairment plays a key role in the development of burn-induced insulin resistance under insulin clamp condition.

*Inhibition of IRS-1/ Akt mediated insulin signaling in skeletal muscle*

We conducted western blot analysis using the gastrocnemius muscles at various time







points after injury to investigate alterations in insulin signaling in skeletal muscle under insulin clamp conditions.    Representative images of blot were shown as Fig.5A. Although IRS-1 expression was not altered by burn injury (**Fig. 5B**), insulin- stimulated tyrosine phosphorylation of IRS-1 in burned mice on PBD 3 was decreased to 34.7% of that in sham-burned mice ($P < 0.05$) (**Fig. 5C**).    Even after normalization by IRS-1 level, insulin-stimulated tyrosine phosphorylation of IRS-1 was also significantly decreased on PBD 3 ($46.9\pm11.7\%$ sham burn, $P < 0.01$) (**Fig. 5D**). As expected, Akt levels were not altered by burn injury (**Fig. 5E**). Insulin-stimulated phosphorylation of Akt was suppressed on PBD 3 ($36.6 \pm 7.7\%$ sham burn, $P < 0.01$) (**Fig. 5E**), and its ratio to Akt was significantly suppressed on PBD 3 ($58.0\pm11.5\%$ sham burn, $P < 0.01$) (**Fig. 5G**). The time-dependent alteration of these insulin signals in skeletal muscle paralled tissue-specific insulin sensitivity assessed byHIC.

**DISCUSSION**

Mice with 30% TBSA full-thickness burn developed whole-body insulin resistance on PBD 3, resulting from impaired insulin-stimulated glucose disappearance from skeletal muscle as determined by a combination study of HIC and microPET, and biochemical analysis of isolated skeletal muscle.    Specifically, there was inhibition of







IRS-1/Akt mediated insulin signaling in parallel with the HIC results.    This is the first report to characterize tissue-specific insulin sensitivity in vivo under insulin clamp conditions following burn injury in a mouse model in association with its molecular alterations.

Initially, we characterized the physiological impact of burn injury in a mouse model and as expected, we found that burn injury induced profound hypermetabolism and catabolism, which were consistent with the clinical course of burn patients with the exception of significant hypoglycemia and hypoinsulinemia following 15-h fasting.    In contrast, models using larger animals such as rabbits and rats developed hyperglycemia similar to that seen with burned patients even after 12-16h of fasting(31, 33), Sugita et al. also reported severe hypoglycemia and hypoinsulinemia after 5-h fasting in a 12% TBSA burned mouse model (26). The fasting blood glucose is primarily determined by the rate of fasting HGP(7), and therefore, these unexpected events in a mouse burn model could be explained by an additive effect of the long fasting time, lack of increased HGP in fasting conditions, and/or extremely high energy expenditure. Energy expenditure is approximately 4 times higher than that of a rat model of 30% TBSA burn injury(32).

Consequently, HOMA-IR, which is a widely accepted index for diabetes derived





from fasting blood glucose and plasma insulin concentrations, led to misinterpretation

of whole-body insulin sensitivity in this burn mouse model on PBD 3 and 7. This

critical finding suggests that use of the surrogate indexes to estimate whole-body insulin

sensitivity under fasting conditions may not be useful for burned mouse models. In the

present study, we also conducted indirect measurements including GTT and ITT to

evaluate whole-body glucose metabolism in the burned animal model up to PBD 7

because previous burn studies(18, 28) had demonstrated significant insulin resistance or

glucose intolerance of burned mice on PBD 3.    Interestingly, the burned mice did not

develop significant glucose intolerance by GTT, which is the most widely used method

for evaluating whole-body glucose tolerance in rodent research. A possible explanation

for this finding is that the peak plasma insulin concentration during GTT was less than

half of that evaluated by HIC in the current study, and the level may have not reached a

concentration high enough to produce differences in insulin-stimulated Rd between

burned and sham-burned mice. Ferrannini et al. also reported that 30-40% of glucose is

taken up by the splanchnic bed during GTT, and HGP is less completely suppressed

than during the insulin clamp technique(8). As expected, ITT demonstrated that burn

injury results in significant whole-body insulin resistance on PBD 3, which resolves by

PBD 7. This drastic time course is consistent with the rapid recovery times of gene







expression (within 4 days), which is the time when the gene changes return back to half after reaching a maximum in a 25% TBSA burn injury mouse model as demonstrated by Seok J at al.(27). These results obtained from the indirect measurements suggest that GTT may not be useful for assessing glucose tolerance in a 30% TBSA burned mouse model.

Next, we conducted parallel HIC and microPET measurements to validate the results of ITT and characterize tissue-specific insulin sensitivity under insulin clamp conditions. HIC showed significant whole-body insulin resistance on PBD 3 in burned mice, which resolved by PBD 7 and was consistent with the ITT results. In addition, HICrevealed that insulin-stimulated peripheral Rd in tissues such as skeletal muscle and/or adipose tissue plays a role in developing the burn-induced insulin resistance. MicroPET following HIC demonstrated that distant skeletal muscle is responsible for impaired insulin-stimulated glucose uptake both visually and quantitatively. Skeletal muscle is responsible for 70 to 80% of whole-body insulin-stimulated glucose uptake under the hyperinsulinemic condition(29) and thus, impaired glucose uptake in skeletal muscle could contribute significantly to whole body insulin resistance in our burned mouse model. In contrast, Xu et al. demonstrated that insulin resistance occurs predominantly in the liver and not in skeletal muscle in a 30% TBSA burned rabbit





model(31). In addition, intensive insulin therapy or treatment with metformin have been reported to suppress HGP and increase the peripheral Rd of burned patients(5, 9, 11). Furthermore, ex vivo studies with burned rodent models have been shown to suppress insulin-stimulated augmentation of 2-deoxyglucose uptake in isolated skeletal muscle(23, 24, 28). The discrepancy of tissue-specific insulin sensitivity among different species emphasizes the need to characterize tissue-specific insulin sensitivity carefully depending on species; otherwise, the results that are obtained may lead to misinterpretation.

Finally, we investigated the molecular alteration in skeletal muscle, which is the tissue responsible for burn-induced insulin resistance in our mouse model. As expected, in gastrocnemius muscle, impaired IRS-1/ Akt mediated insulin signal transduction, which is related to the time course of reduced insulin sensitivity in vivo obtained with HIC. Most of the previous molecular investigations of insulin signal transduction were conducted after various doses of insulin challenge in which the plasma insulin concentration and blood glucose level of the animal models were not clearly documented. Cree et al. successfully demonstrated that peroxisome proliferator-activated receptor gamma treatment could improve skeletal muscle insulin sensitivity in children with burns involving greater than 40% TBSA at 3 weeks







post-burn using insulin clamp techniques(5). This was also associated with improvement in insulin-stimulated IRS-1 expression in skeletal muscle.    We believe that it is important to investigate the molecular basis of burn-induced insulin resistance and the effects of novel treatments with consideration of insulin and glucose state. Our burned mouse model could provide a more accurate molecular and genetic basis of possible treatments including insulin intensive therapy(9)(30), beta-blockade(12), farnesyltransferase inhibitor(23), and mitochondria-targeted antioxidant peptides(18).

## CONCLUSIONS

This study clearly demonstrated that impaired glucose utilization in skeletal muscle, not elevated hepatic glucose production, plays a pivital role in the development of burn-induced insulin resistance in a mouse model. This burned mouse model could contribute to further explorations of the molecular and genetic alteration associated with burn-induced insulin resistance and may set the stage for developing more targeted approaches to treat altered glucose metabolism in burned patients.

## ACKNOWLEDGEMENTS

The authors thank Florence Lin for her great technical support.





## GRANTS

This work was supported by grants from National Institute of Health (P50-21700;

P30DK0040561) and from Shriners Medical Research SHCC #84070 and #85500.

## DISCLOSURES

No conflicts of interest, financial or otherwise, are declared by the authors.

## AUTHOR CONTRIBUTIONS

R.G.T., A.J.F., M.K., and Y.M.Y. conception and design of research

T.Y., Y.I., V.V.B., and Y.M.Y. performed experiments

T.Y. and Y.M.Y. analyzed data

T.Y., Y.M.Y., Y.I., V.V.B., M.K., M.I.P., A.J.F., and R.G.T. interpreted results of

experiments

T.Y. prepared figures

T.Y. and Y.M.Y. drafted manuscript

T.Y., Y.M.Y., Y.I., V.V.B., M.I.P., S.I, M.F.W., A.J.F., and R.G.T. edited and revised

manuscript

R.G.T. approved final version of manuscript

**FIGURE LEGENDS**

**Figure 1. Burn-induced hypermetabolism and catabolism in a mouce model**

Dynamic change in body weight (A), organ weight of gastrocnemius muscle,

epididymal adipose tissue and liver (B), 24-h energy expenditure (C-E), oxygen

consumption ($VO_2$) (F), and resting carbon dioxide production ($VCO_2$) (G) on

post-burn day (PBD) 1, PBD 3, and PBD 7 in burn and sham-burn pair-fed (SBP) mice

at various time points (n=5-8 mice per group).    Data are expressed as mean ± SEM. *P

< 0.05; **P < 0.01 vs SBP.

**Figure 2. Systemic insulin resistance emerged on post-burn day 3 by indirect**

**measurements**

Blood glucose level (A-C) and plasma insulin concentration (D-F) during

intraperitoneal glucose tolerance test (2 g/kg BW) on post-burn day (PBD) 1, 3, and 7 in

burn and sham-burn pair-fed (SBP) mice (n=8 mice per group). Intraperitoneal insulin

tolerance test (1 U/kg BW) on PBD 1, 3, and 7 (n=8 per group) (G-I). Data are

expressed as mean ± SEM. *P < 0.05 vs SBP.

**Figure 3. Burned mice decreased insulin-stimulated glucose utilization in the**







**peripheral tissues by hyperinsulinemic isoglycemic clamp**

Hyperinsulinemic isoglycemic clamp results for blood glucose concentration(A-C), glucose infusion rate (GIR) (D-F), correlation between GIR and index of insulin sensitivity (SI$_{clamp}$) (G), Hepatic glucose production (HGP) (H), rate of glucose disappearance (Rd) (I), , plasma insulin concentration (J) in burn and sham-burn pair-fed (SBP) mice at post-burn day (PBD) 1,3, and 7 (n=5-7 mice per group). Data are expressed as mean ± SEM. *P < 0.05 and **P < 0.01 vs SBP.

**Figure 4. Burned mice decreased insulin-stimulated glucose utilization in the skeletal muscle by hyperinsulinemic isoglycemic clamp and microPET**

Typical images of whole body microPET at basal and clamp period (insulin infusion rate of 2.5 mU/kg/min) in burn and sham burn pair-fed (SBP) mice at post-burn day 3 (n=4 mice per group) (A). White arrow indicates the quadriceps. Standardized uptake value (SUV) in quadriceps (B). Data are expressed as mean ± SEM. *P < 0.05; #P < 0.01 vs SBP clamp.

**Figure 5. Burn injury inhibited IRS-1 and Akt mediated insulin signaling in the skeletal muscle by immunoblotting**







Typical images of immunoblotting (A), IRS-1(B), tyrosine phosphorylated IRS-1 (C),

and insulin-stimulated phosphorylation IRS-1 ratio normalized to IRS-1 (D), Akt (E),

serine phosphorylated Akt (F), and insulin-stimulated phospholiration Akt normalized to

Akt in burn and sham burn pair-fed (SBP) mice on post burn day (PDB) 1, 3, and 7 (G).

Data are expressed as mean ± SEM. *P < 0.05 and **P < 0.01 vs SBP.






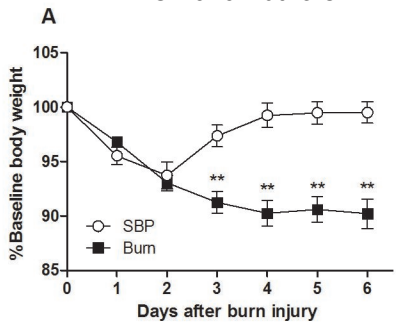

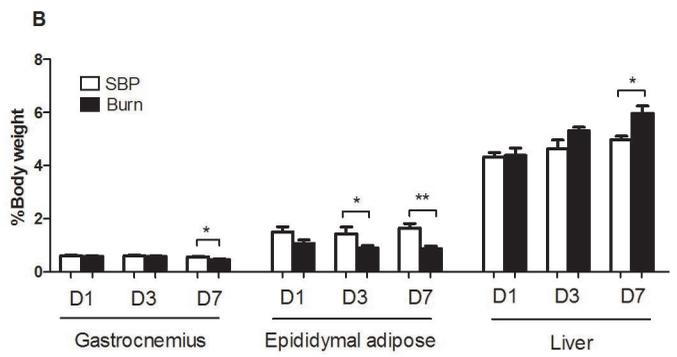

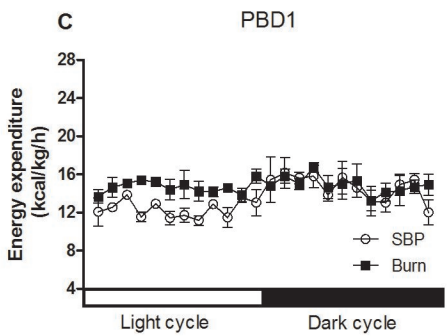

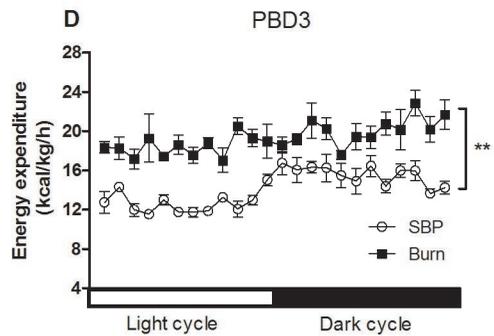

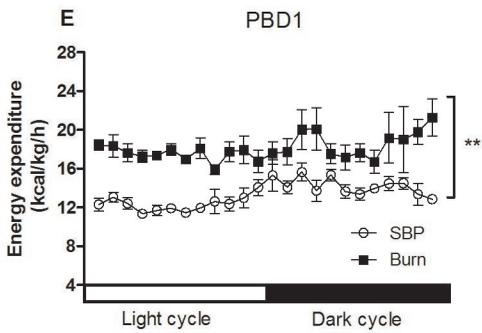

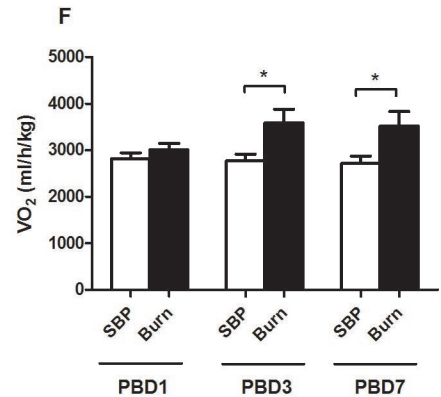

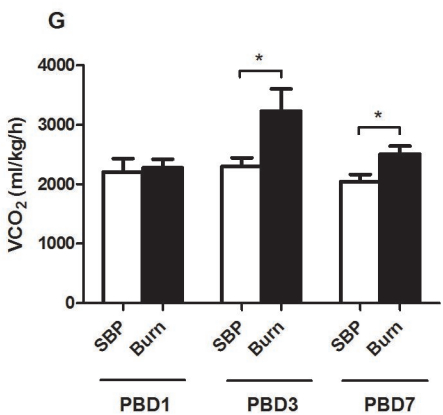





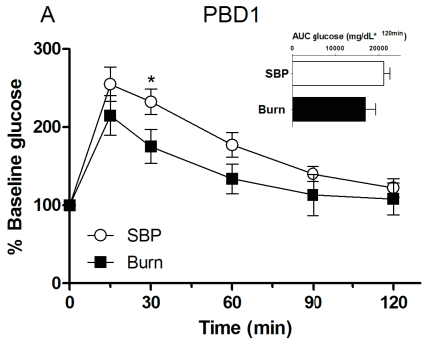

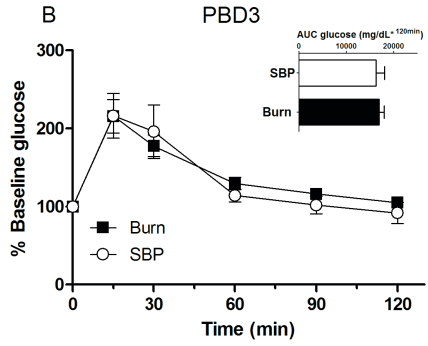

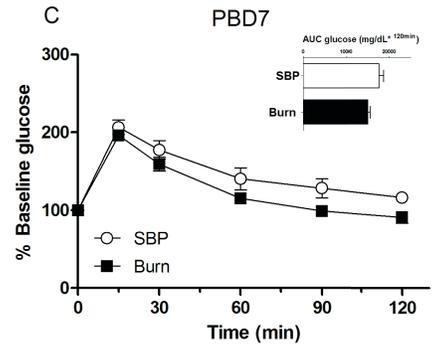

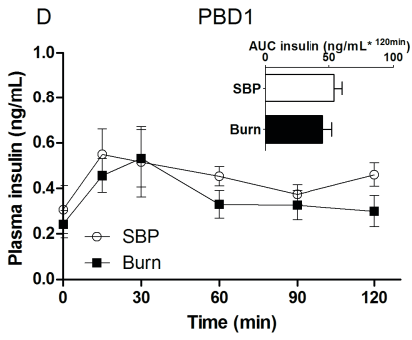

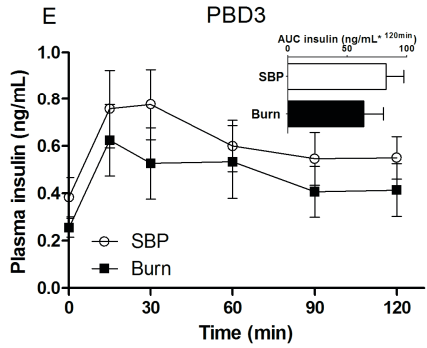

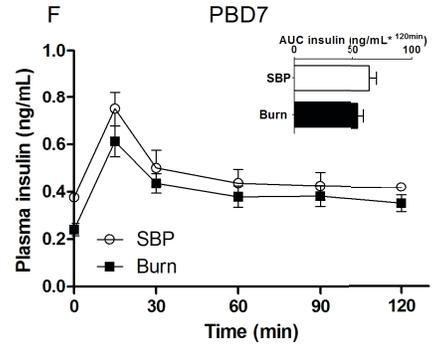

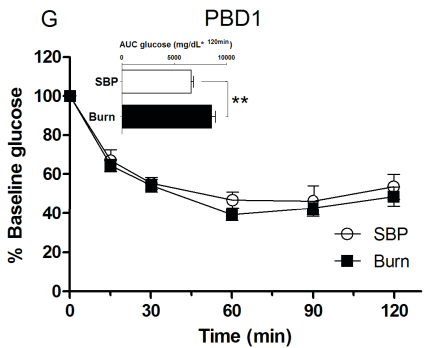

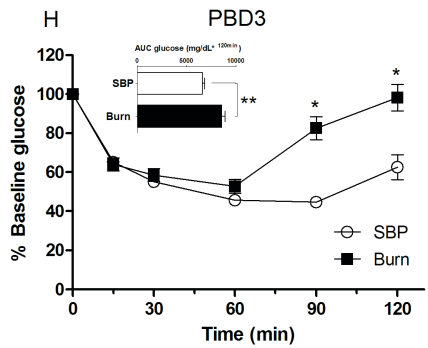

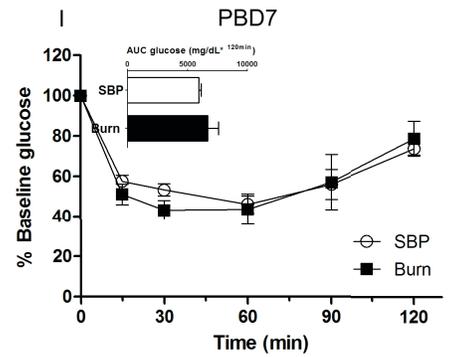



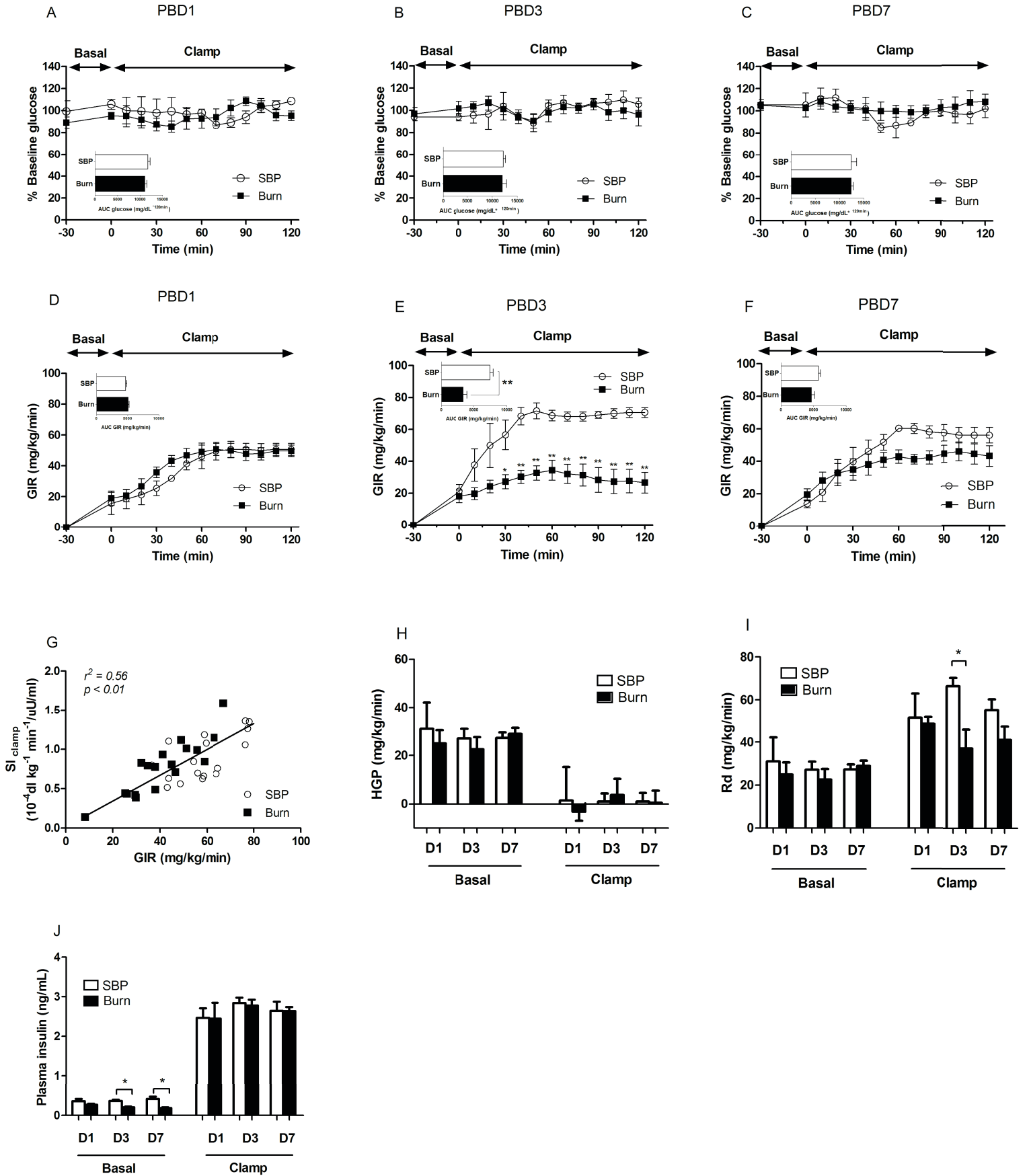





A

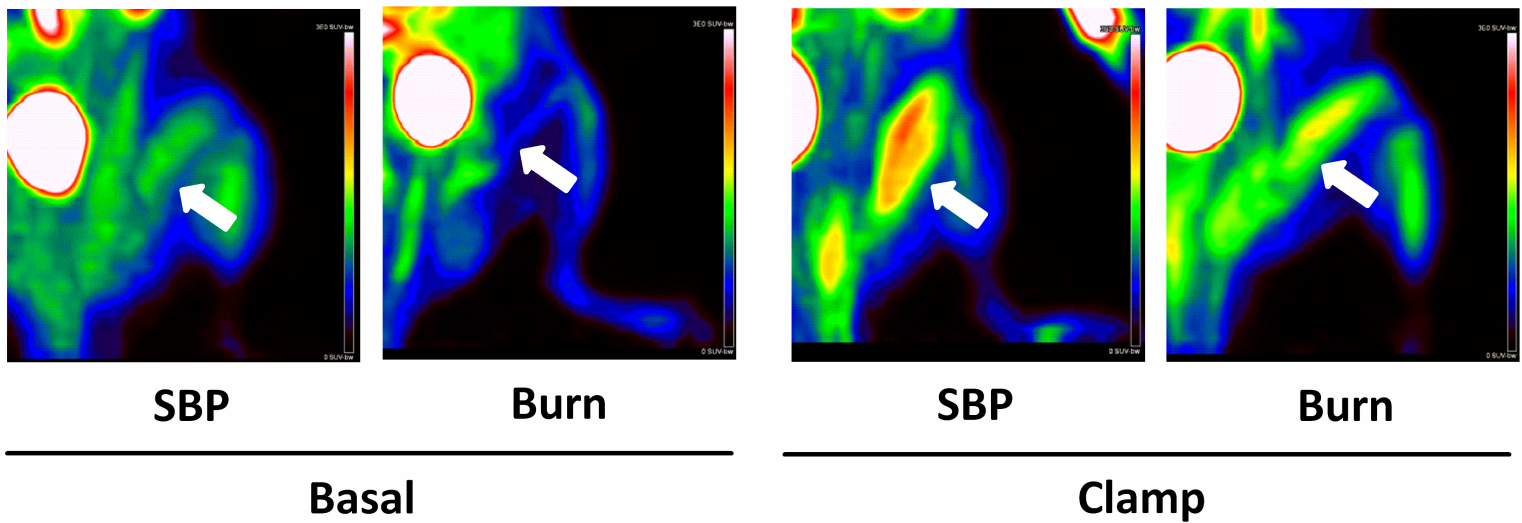

B

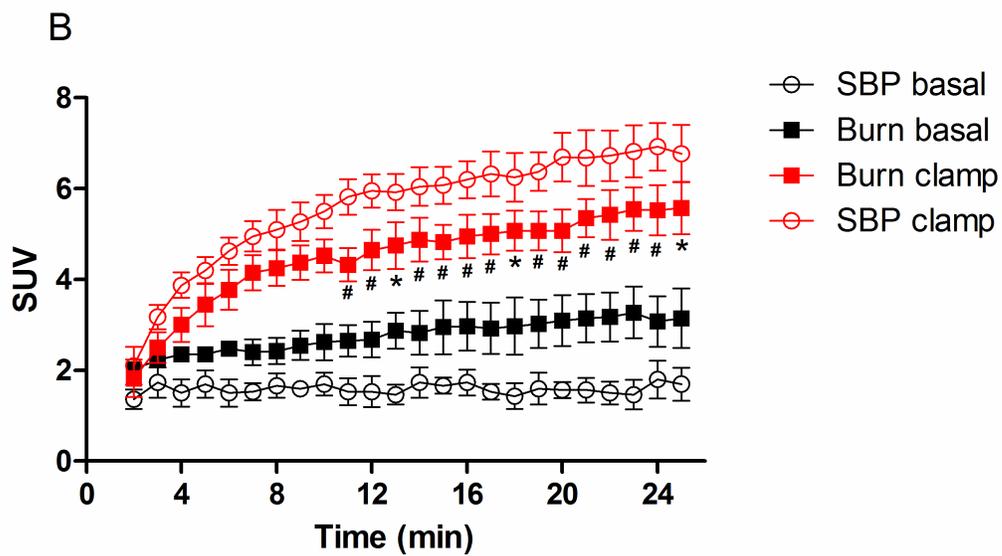









**Table 1. Dynamic change of blood glucose and plasma insulin concentration after burn injury**

| Post-burn day | 1 | 3 | 7 |
|---|---|---|---|
| **Blood glucose concentration (mg/dl)** | | | |
| Sham burn | 77.7 ± 4.4 | 87.3 ± 5.3 | 81.5 ± 6.6 |
| Burn | 67.7 ± 6.0 | 68.5 ± 4.1** | 67.3 ± 6.5* |
| **Plasma insulin concentration (μU/ml)** | | | |
| Sham burn | 8.8 ± 0.7 | 9.0 ± 1.2 | 10.2 ± 1.8 |
| Burn | 8.1 ± 1.1 | 5.6 ± 0.7** | 5.8 ± 1.0** |
| **HOMA-IR** | | | |
| Sham burn | 1.8 ± 0.2 | 2.0 ± 0.4 | 2.1 ± 0.4 |
| Burn | 1.2 ± 0.4 | 1.0 ± 0.1** | 0.9 ± 0.2** |

Blood sampling was conducted in burn and sham burn mice after 15 h fasting. Homeostatic model assessment of insulin resistance (HOMA-IR) was calculated according to the formula: fasting blood glucose (mg/dl) $\times$ fasting plasma insulin concentration (μU/ml) $\times$ $405^{-1}$. Data are expresed as mean ± SEM. *$P < 0.05$; **$P < 0.01$ vs Sham burn. n=5-8 per each group.